# Aging, rejuvenation and thixotropy
# in complex fluids.

## Time-dependence of the viscosity
## at rest and under constant shear rate or shear stress[1].

D. Quemada

*Laboratoire Matière et Systèmes Complexes- UMR CNRS 7057.*
*Université Paris Diderot-Paris7, bâtiment Condorcet,*
*Mail address: MSC, case 7056, Univ.Paris7, 75205 Paris-cedex 13*
<danielquemada@orange.fr>

**Abstract** Complex fluids exhibit time-dependent changes in viscosity that have been ascribed to both thixotropy and aging. However, there is no consensus for which phenomenon is the origin of which changes. A novel thixotropic model is defined that incorporates aging. Conditions under which viscosity changes are due to thixotropy and aging are unambiguously defined. Viscosity changes in a complex fluid during a period of rest after destructuring exhibit a bifurcation at a critical volume fraction $\phi_{c2}$. For volume fractions less than $\phi_{c2}$ the viscosity remains finite in the limit $t \to \infty$. For volume fractions above critical the viscosity grows without limit, so aging occurs at rest. At constant shear rate there is no bifurcation, whereas under constant shear stress the model predicts a new bifurcation in the viscosity at a critical stress $\sigma_B$, identical to the yield stress $\sigma_y$ observed under steady conditions. The divergence of the viscosity for $\sigma \leq \sigma_B$ is best defined as aging. However, for $\sigma > \sigma_B$, where the viscosity remains finite, it seems preferable to use the concepts of restructuring and destructuring, rather than aging and rejuvenation. Nevertheless, when a stress $\sigma_A (\leq \sigma_B)$ is applied during aging, slower aging is predicted and discussed as true rejuvenation. Plastic behaviour is predicted under steady conditions when $\sigma > \sigma_B$. The Herschel-Bulkley model fits the flow curve for stresses close to $\sigma_B$, whereas the Bingham model gives a better fit for $\sigma >> \sigma_B$. Finally, the model's predictions are shown to be consistent with experimental data from the literature for the transient behaviour of laponite gels.

**Résumé :** Un modèle thixotrope qui décrit la viscosité $\eta(t)$ au cours d'une période de repos, consécutive à une déstructuration, prédit l'existence d'une bifurcation pour une fraction volumique critique $\phi_{c2}$. Pour $\phi < \phi_{c2}$, la limite $\eta(t \to \infty)$ reste finie tandis que pour $\phi \geq \phi_{c2}$, la viscosité croît sans limite. C'est dans ce second domaine qu'a lieu le vieillissement. A l'inverse de l'absence de bifurcation lorsque le système est sous vitesse de cisaillement constante, le même modèle prédit l'existence, sous contrainte constante $\sigma$, d'une nouvelle bifurcation de $\eta(t)$ pour une contrainte critique $\sigma_B$ qui s'identifie au seuil de contrainte $\sigma_y$ observé en régime stationnaire. La divergence de $\eta(t)$ lorsque $\sigma \leq \sigma_B$ est de nouveau associée au vieillissement, mais il semble préférable d'utiliser les concepts de restructuration et de déstructuration plutôt que de vieillissement et de rajeunissement dans le domaine $\sigma > \sigma_B$ où la viscosité reste finie. Néanmoins, lorsqu'on applique une contrainte $\sigma_A$ à un système en cours de vieillissement, le modèle prédit (si $\sigma_A \leq \sigma_B$) un *ralentissement du vieillissement* qui semble pouvoir être considéré comme un véritable rajeunissement. De plus, le comportement plastique, qui est prédit en régime stationnaire dans le domaine $\sigma > \sigma_B$, est conforme au modèle d'Herschel-Bulkley au voisinage de $\sigma_B$ mais à celui de Bingham pour $\sigma >> \sigma_B$. Finalement, les prédictions du modèle sous vitesse de cisaillement constante, sont discutées par comparaison avec des mesures de vieillissement et de rajeunissement dans des gels de laponite.

**Mots-clés : transition pâteuse, modèle thixotrope, bifurcation, seuil de contrainte, vieillissement, rajeunissement.**

---

[1] Most of this paper has been previously published in French [Quemada, 2004]

## 1. Introduction.

Recent studies of complex systems have shown strong analogies between the rheological behaviour of pastes (such as concentrated suspensions and emulsions, gels, foams,…) and glasses (such as amorphous solids and spin glasses) [**Larson, 1999 ; Cloitre et al, 2000**]. The liquid-to-paste transition is associated with the emergence of a yield stress, $\sigma_y$, as the volume fraction, $\phi$, increases. Under steady conditions, as soon as $\sigma > \sigma_y$, *non-linear plastic* behaviour appears. It can often be described by the Herschel-Bulkley (HB) model: $\sigma = \sigma_y + K \dot{\gamma}^n$ (see [**Cloitre et al, 2000**], for instance). This yield stress has been considered as revealing a bifurcation in the viscosity of the system.

Coussot *et al.* observed a bifurcation in the time evolution of the viscosity in three very different complex systems (an aqueous suspension of bentonite, a polymeric gel and a colloidal glass of laponite), under constant shear stress [**Coussot et al, 2002a & 2002b**]. All three systems exhibited a critical bifurcation stress, $\sigma_B$. When $\sigma > \sigma_B$ the limiting viscosity $\eta(t \rightarrow \infty)$ tended towards a finite value, whereas when $\sigma < \sigma_B$ $\eta \rightarrow \infty$ as $t \rightarrow \infty$.

Analogies between pastes and glasses have been attributed to two shared characteristics: *structural disorder* and *metastability*. Their origin is considered to be a dynamics in which thermal energy alone is insufficient to cause complete structural relaxation. Given these analogies, most of the modelling has been based on adding a rheological component to classical models for the glass transition in liquids and spin glasses [**Bouchaud et al, 1995; Mason & Weitz, 1995 ; Sollich et al, 1997 ; Hebraud & Lequeux, 1998**]. In concentrated dispersions and gels, a *paste transition* corresponds to the glass transition in classical solids and will represent the "jamming transition" observed in these systems [**Trappe et al, 2001 ; Segré et al, 2001**]. The main measurable characteristics of pastes described by these models are the time evolution of the viscosity and the relaxation modulus [**Abou at al, 2003 ; Derec et al, 2001**].

These time-dependent viscosity changes are very similar, if not identical, to what rheologists have long defined as "thixotropy" [**Mewis, 1979 ; Barnes, 1997**]. Nevertheless, in the recent literature, what would previously have been called thixotropic behaviour has often been interpreted in terms of "aging" and "rejuvenation". In addition, no attempt has been made to reconcile these two paradigms for viscosity changes. Thus, the current situation is ambiguous: are thixotropy and aging/rejuvenation identical? If not, under what circumstances is each the appropriate description? The model defined here is a contribution to resolving this ambiguity, since the two phenomena coexist and have separate, well-defined roles.

## 2. The non-linear structural (NLS) model

The complex system is approximated as a polydisperse dispersion, composed of individual particles (IP) which form structural units (SU) at high volume fraction *(reference?)*. IP and SU are considered in the framework of the hard sphere (HS) approximation. They have radii of $a$ and $a_{eff}$, respectively.

The model is based on the following $\eta(\phi)$-relation for hard spheres: $\eta = \eta_F (1 - \phi/\phi_m)^{-2}$, with $\eta_F$ the viscosity of the suspending fluid and $\phi_m$ the maximum packing fraction. This equation has been widely used for HS suspensions in the literature [**de Kruif *et al*, 1985; Brady, 1993; Rueb and Zukoski, 1998; Heyes and Sigurgeirsson, 2004**]. Here, it is generalized for the viscosity of complex fluids (see [**Quemada, 1998**] for more details) to give:

$$\eta = \eta_F (1 - \phi_{eff} /\phi_m)^{-2} \tag{1}$$

with $\phi_{eff}$ the effective volume fraction, which depends on the degree of structuring via the equation:

$$\phi_{eff} = (1 + CS) \, \phi \tag{2}$$

where $S$ is the fraction of individual particles included in structural units and $C=(\varphi^{-1} - 1)$ is a factor related to the average SU compactness $\varphi$. The fraction $S$ is the *structural variable* of the model. The limiting viscosities at low and high shear, $\eta_0$ and $\eta_\infty$, correspond respectively to complete structuring ($S=1$, under very low shear or at rest) and complete destructuring ($S=0$, under very high shear). They are given by:

$$\eta_0 = \eta_F (1 - \phi /\phi_0)^{-2} \tag{3}$$

$$\eta_\infty = \eta_F (1 - \phi /\phi_\infty)^{-2} \tag{4}$$

where

$$\phi_0 = \varphi \, \phi_m \quad \text{and} \quad \phi_\infty = \phi_m \tag{5}$$

are the maximum effective packing fractions in these two limits (see also the end of section 6).

Under constant shear stress, $\sigma$, or shear rate, $\dot{\gamma}$, the time evolution of the system (SU+IP) is described by a kinetic equation for $S$. For simplicity, it is assumed to be a relaxation equation, given by:

$$\frac{dS}{dt} = \kappa_F ( 1 - S ) - \kappa_D S \qquad\qquad (6)$$

The kinetic constants for formation, $\kappa_F$, and erosion, $\kappa_D$, of SU depend on a reduced shear rate or shear stress, $\Gamma = (\sigma/\sigma_C)$ or $(\dot{\gamma}/\dot{\gamma}_C)$, where $\sigma_C$ and $\dot{\gamma}_C$ are critical values characteristic of the particular system. Inserting the solution of eq. (6), $S(\Gamma, t)$, into eq.(1) automatically gives $\eta(\Gamma, t)$. Thus, we can model both the non-Newtonian and thixotropic properties of complex fluids, as can most structural models from the literature. However, note that the latter generally (and empirically) use a *linear* relation $\eta(S)$, whereas the relation is non-linear in the present model. For this reason, we call it the *Non-Linear Structural (NLS) model*. In addition, the non-linearity is not empirical, but based on the physics underlying the viscosity relation, eq.(1), see Quemada, 1998.

In dilute systems, non-Newtonian properties result from the competition between Brownian diffusion of particles and the friction exerted by the suspending fluid. These two forces respectively create and destroy SU. Taking in eq.(6) $\kappa_F \propto t_{Br}^{-1}$ and $\kappa_D \propto \dot{\gamma}$, this competition depends on the ratio $\kappa_D/\kappa_F$, *i.e.* the Péclet number, $Pe = t_{Br}\dot{\gamma} = 6\pi\eta_F\, a^3\, \dot{\gamma}/KT$ where $t_{Br} \approx a^2/D_0$ is the Brownian diffusion time in the continuous phase. $D_0 \approx KT/a\eta_F$ is the diffusion coefficient in this phase.

As the volume fraction of particles increases, the diffusion of particles is increasingly hindered by their neighbours. This effect can be approximated by assuming that each particle diffuses through an effective medium with the viscosity of the dispersion. Then the diffusion time $t_{Diff} \approx \kappa_F^{-1} \approx \eta\, a^3 / KT$ appears as a variable which depends on the structure through $\eta = \eta(S)$. With $\kappa_D/\kappa_F \approx \eta\dot{\gamma}$ (in the absence of any interaction potential), one thus obtains an effective Péclet number:

$$Pe^* \approx \sigma\, a^3/KT = \sigma/\sigma_{CR} \qquad\qquad (7)$$

with $\sigma_{CR} \approx KT / a^3$ a critical stress. Hence, from eq.(1), $t_{Diff}$ is an increasing function of $\phi$, due to increased caging of particles, which increases both the probability of collision and the hydrodynamic interactions. This point is confirmed by the decrease, as $\phi$ is increased, of the two limits of the short and long time self-diffusion coefficients, $D_S{}^S$ and $D_S{}^L$, which represent local particle motions at distances respectively less than $a_{eff}$ and greater than a few $a_{eff}$. Therefore one obtains a structure-dependent slow relaxation mode, as the self-diffusion coefficients, specially the long-time one [**Morris and Brady, 1996**]. As the glass transition (defined by $\phi_g$) is approached, particle diffusion slows. When $\phi_{eff} = \phi_g$ ($D_S{}^L \to 0$) the particles are totally confined inside the transient cages formed by their neighbours. On the other hand, at volume fractions above $\phi_g$, $D_S{}^S$ remains finite, since particles can still vibrate within their cages. As the volume fraction increases, the cage size decreases, so this motion decreases. The limit $D_S{}^S \to 0$ is reached when the volume fraction reaches random close packing (RCP) $\phi_{eff} = \phi_m \equiv \phi_{RCP}$ [**Brady, 1993; Knaebel et al, 2000**]. This last limit agrees with the divergence of the zero shear viscosity, which has often been observed at this concentration [**Heyes and Sigurgeirsson, 2004**]. From these arguments, it seems quite justified to consider that the range $\phi_g < \phi_{eff} < \phi_m$ is the domain of the paste phase, where particle movement is restricted, but not impossible.

## 3 – The paste transition: Time evolution of viscosity at rest and aging.

### 3.1 – Effect of volume fraction on $\phi_{eff}(t)$

The time-evolution of the viscosity is now analysed for a system at rest after destructuring (for instance, by filtering or application of a large amplitude oscillation). With the time-dependent viscosity $\eta[\phi_{eff}(t)]$, the kinetic constant of formation can be written $\kappa_F \approx (\eta\, a^3/KT)^{-1} = \kappa_{A0}/\eta_R(t) = \kappa_F(t)$, where $\kappa_{A0} = (6\pi\eta_F\, a^3/KT)^{-1}$, is a constant with units of $s^{-1}$. A natural unit for the dimensionless time is $\kappa_{A0}t$. As the viscosity $\eta(t)$ of the system during restructuring is given by eq.(1), we obtain:

$$\kappa_F = \kappa_{A0}(1 - \phi_{eff}/\phi_m)^2 \qquad (8)$$

At rest, $\kappa_D \approx \dot{\gamma} = 0$ and eq.(3) becomes $dS/dt = \kappa_F(1-S)$. With eq.(8), this equation can be solved iteratively, starting from $S = S_{init}$.

Based on these considerations, Fig.1 shows typical results for the time-evolution of the effective volume fraction, $\phi_{eff}(t)$, as a function of dimensionless time. In particular, it shows how the final state of the system depends on the true volume fraction $\phi$. Note that, as discussed at the end of the previous section, the two horizontal straight lines $\phi_{eff} = \phi_g$ and $\phi_{eff} = \phi_m$ limit two domains with different behaviour : fluid for $\phi_{eff} < \phi_g$ and paste for $\phi_g < \phi_{eff} < \phi_m$. A critical value $\phi = \phi_{c2}$ divides the paste domain in two, depending on the limit of $\phi_{eff}$ as $t \to \infty$: $\phi_{eff}(\infty)$ remains less than $\phi_m$ when $\phi < \phi_{c2}$, while $\phi_{eff}(\infty) \to \phi_m$ when $\phi \geq \phi_{c2}$. Thus two states of the system can be distinguished: 1) when the steady viscosity remains finite, the system will be called a "*soft paste*"; 2) when the steady viscosity tends to infinity the system will be considered a "*hard paste*".

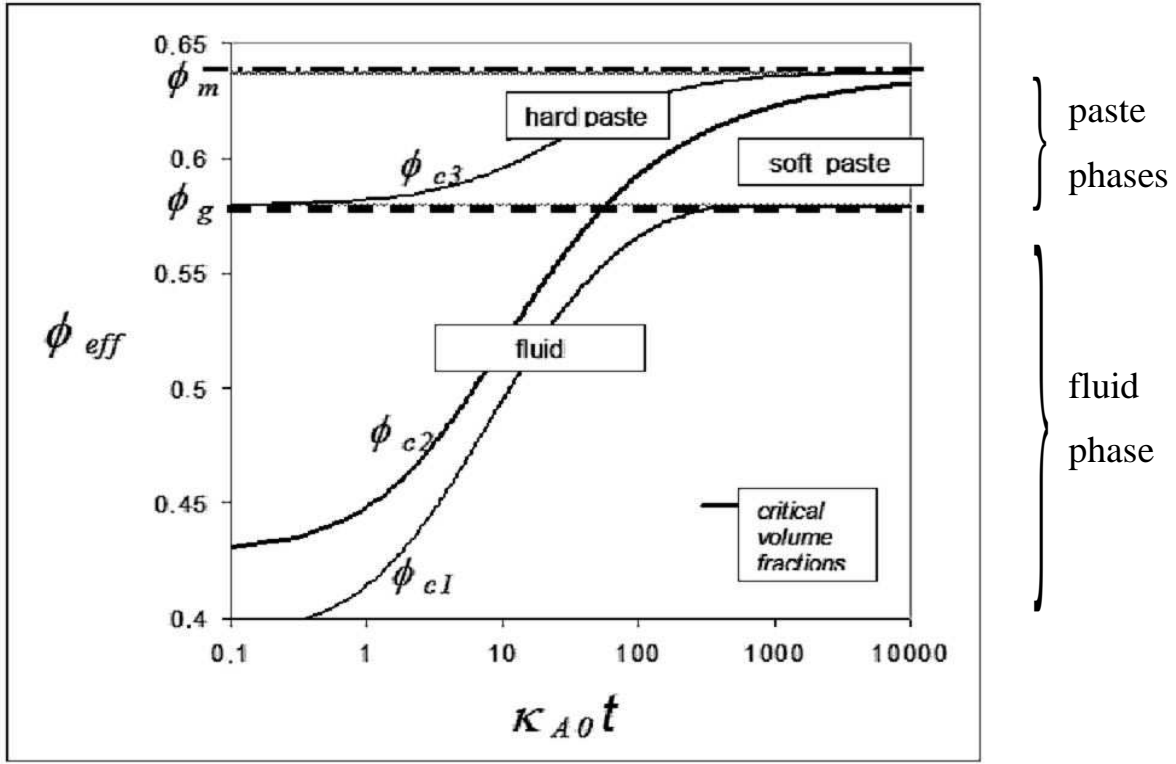

**Fig.1 – Evolution of the effective volume fraction vs. the reduced time $\kappa_{A0}\,t$. The curves $\phi_{c1}$, $\phi_{c2}$ and $\phi_{c3}$ divide the domain into sub-domains corresponding to the different states of the system.** NLS-model parameters : $\eta_F = 1\ mPas;\ \phi_m = \varphi = 0.637;\ S_{init} = 0.1;\ \kappa_{A0} = 10^3\ s^{-1}$ )

There are two other critical volume fractions. First, $\phi = \phi_{c1}$ is the lowest volume fraction for which the glass phase can be reached: $\phi_{eff}(\infty) \to \phi_g$ and second, $\phi = \phi_{c3}$ is the lowest volume fraction for which the system is in the glass phase at time zero: $\phi_{eff}(0) = \phi_g$. If $\phi < \phi_{c1}$, the system will remain fluid; if $\phi > \phi_{c3}$, the system is initially in the hard paste state and remains so indefinitely. For complete restructuring, i.e. when $S = 1$ at $t \to \infty$, we can use eq.(2) to obtain: $\phi_{c1} = \varphi\phi_g$, $\phi_{c2} = \varphi\phi_m$ and $\phi_{c3} = \phi_g\,/(1 + CS_{init})$.

For the example shown in Fig.1, the critical values are $\phi_{c1} = 0.369$, $\phi_{c2} = 0.406$ and $\phi_{c3} = 0.548$. These values are due to the choice of $\varphi = 0.637$ for the mean compactness of SU. This choice does not affect the shape of the state diagram. It only influences the position of the critical values. Table 1 summarises the behaviour corresponding to the various ranges of volume fraction.

***Table 1. Different types of behaviour***

| Domain | Behaviour | Initial and final effective volume fractions |
|---|---|---|
| $\phi < \phi_{c1}$ | Fluid | $\phi_{eff}(0) < \phi_g$ ; $\phi_{eff}(\infty) < \phi_g$ |
| $\phi_{c1} < \phi < \phi_{c2}$ | Fluid → Soft Paste | $\phi_{eff}(0) < \phi_g$ ; $\phi_{eff}(\infty) < \phi_{RCP}$ |
| $\phi_{c2} < \phi < \phi_{c3}$ | Fluid → Hard Paste | $\phi_{eff}(0) < \phi_g$ ; $\phi_{eff}(\infty) = \phi_{RCP}$ |
| $\phi_{c3} < \phi < \phi_{RCP}$ | Hard Paste | $\phi_{eff}(0 > \phi_g$ ; $\phi_{eff}(\infty) = \phi_{RCP}$ |

### 3.2 – Effect of volume fraction on $\eta(t)$.

Fig.2 shows the time evolution of the relative viscosity corresponding to these different kinds of behaviour at different, constant volume fractions.

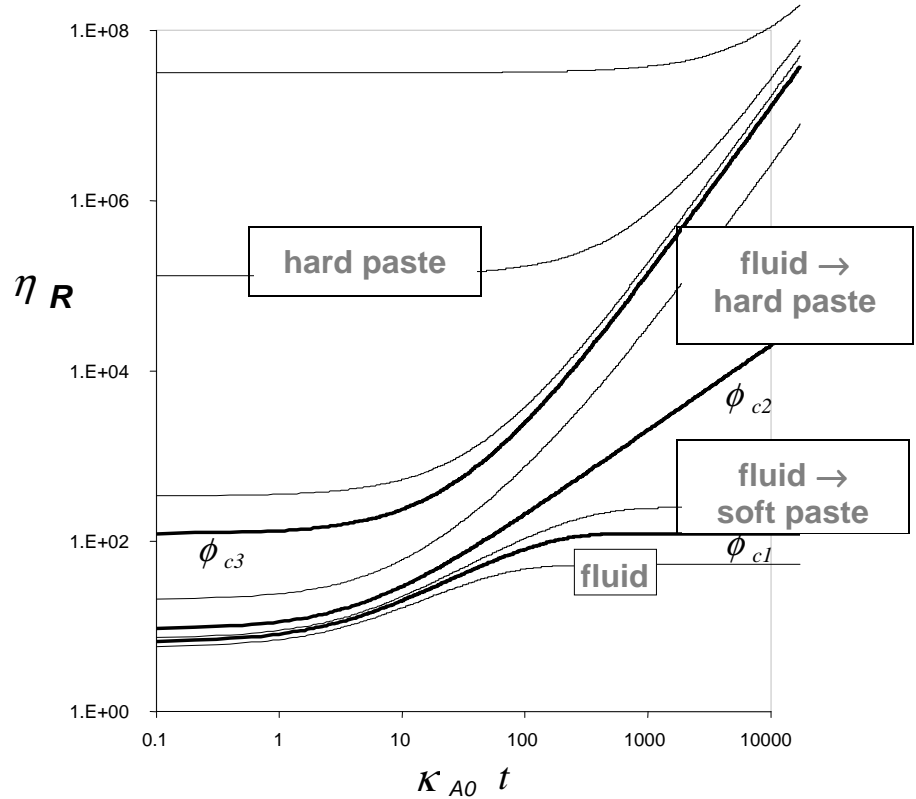

***Fig.2 – Relative viscosity vs. reduced time $\kappa_{A0}\, t$. The bold curves $\phi_{c1}$, $\phi_{c2}$ et $\phi_{c3}$ separate the different domains of behaviour shown in Table 1. From bottom to top, the six intermediate curves correspond to $\phi = 0.35$; 0.38 ; 0.47 ; 0.5; 0.601 ; 0.6025. The top two curves, with $\phi > \phi_{c3}$, show the extremely rapid variation of the initial value of $\eta$ as $\phi$ approaches $\phi_m$. Same values of the NLS-model parameters were used as for Fig.1.***

Studying the long time evolution of $\eta(t)$ allows us to answer the following question, at least partly: Can the viscosity changes of hard pastes be regarded as aging?

Starting from the expressions for $\eta$, $\phi_{eff}$, $dS/dt$ and $\kappa_F(t)$, and putting

$$\alpha = \phi/\phi_{c2} - 1 \qquad (9)$$

one obtains the following differential equation for the function $u = (1 - \phi_{eff}/\phi_m)^{-1}$

$$\frac{du}{\alpha + u^{-1}} = \kappa_{A0} dt \qquad (10)$$

Crossing the boundary $\phi = \phi_{c2}$ changes the sign of $\alpha$. This leads to three types of solution to eq.(10), hence to three different types of behaviour, depending on whether $\alpha < 0$, $\alpha > 0$ or $\alpha = 0$.

A) If $\alpha < 0$, $\phi_{eff}(t \to \infty) < \phi_m$, so the limit of $u$ is finite, and hence for $\eta$:

$$\eta(t \to \infty) \approx \eta_F \alpha^{-2}\{1 - exp[-\alpha^2 \kappa_{A0} t]\}^2 \qquad (11)$$

which corresponds well to the viscosity limit at zero shear rate, eq.(3), with zero shear packing $\phi_0 = \varphi \phi_m$, shown within the domain $\phi < \phi_{c2}$ in Fig.2.

B) If $\alpha > 0$, one has $\phi_{eff} \to \phi_m$ if $t \to \infty$, so both $u$ and $\eta$ tend to infinity. Hence, the long time solution of eq.(10) reduces to $u \approx \alpha \kappa_{A0} t$, giving:

$$\eta(t \to \infty) \approx \eta_F \alpha^2 (\kappa_{A0} t)^2 \qquad (12)$$

This result agrees with the asymptotic limit of the curves in Fig.2 for $\phi > \phi_{c2}$: a straight line with slope two in logarithmic coordinates[2].

C) If $\alpha = 0$, eq.(10) gives $u^2 = 2\kappa_{A0}t + const$ and, putting $\eta_{init} = \eta(t=0)$, leading to:

$$\eta = \eta_{init} + 2\eta_F \kappa_{A0} t \qquad (13)$$

for the whole of time dependence, agreeing with the asymptotic limit of $\eta$ along the boundary $\phi = \phi_{c2}$ in Fig.2: a straight line with slope one in logarithmic coordinates.

Therefore, the rheological behaviour of the system at rest shows a *bifurcation* at the critical volume fraction $\phi = \phi_{c2}$. Moreover, at rest, it appears that the absence of a limiting viscosity when $\phi \geq \phi_{c2}$ is due to both aging and metastability of the system, as shown in the next section.

---

[2]  Clearly, this quadratic asymptotic behaviour is due to the form of eq.(1) with an exponent of $-2$. Changing to an exponent of $-q$ will lead automatically to an asymptotic behaviour in $t^q$.

### 3.3 – Long time behaviour of hard pastes: aging and metastability

In the framework of the analogy between the paste and glass transitions, the behaviour described by eq.(12) when $\phi > \phi_{c2}$ seems to correspond to the lack of equilibrium observed in glassy phases below the glass transition temperature $T_g$. Moreover, one also recovers the fact that the characteristic time $t_{Str}$ for "structuring", associated to the evolution of $\eta(t)$, becomes longer and longer as time elapses. From eq.(12), using $t_{Str} \approx \eta(d\eta/dt)^{-1}$ to estimate the order of magnitude of this time, one obtains

$$t_{Str} \approx t \qquad (14)$$

This dependence agrees with stress relaxation measurements of colloidal gels submitted to oscillations of very weak amplitude after some rest period of duration $t_w$: these measurements show that the order of magnitude of the relaxation time evolves as $t_w$, *i.e.* as the *age* of the system, which is a general characteristic of physical aging (see [**Knaebel** *et al*, **2000**], for instance). As this age is identical to the evolution time of $\eta(t)$, it seems justifiable to speak about *aging* when $\phi > \phi_{c2}$ .

Furthermore, as expected, restructuring is slower for harder pastes, *i.e.* when the initial effective volume fraction (at the end of the preparation) is closer to $\phi_m$. In other words, the fluid-paste transition (at $\phi_{eff} = \phi_g$) occurs sooner, leading to a slower restructuring.

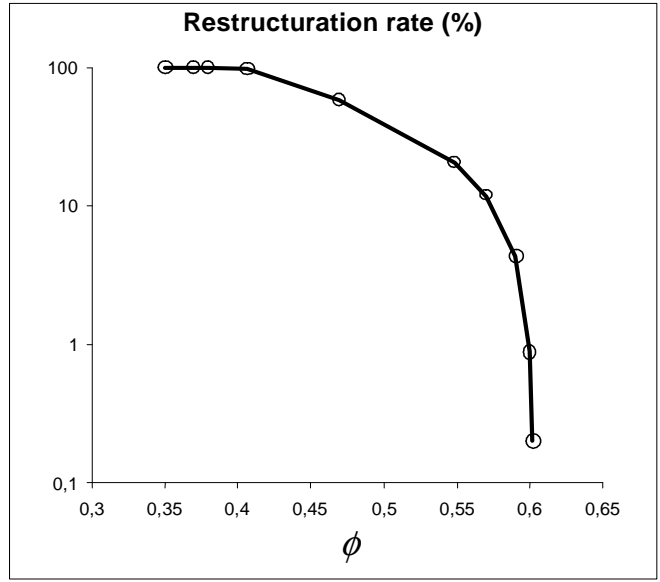

**Fig.3 Rate of restructuring as a function of $\phi$. The fall is more and more abrupt as $\phi$ increases above $\phi_{c2}$ (here $\phi_{c2} = 0.406$) and becomes extremely steep as $\phi \to \phi_m$. Same values of NLS-model parameters as in Fig.1, with $\kappa_{A0} t_{fin} \approx 2.10^{-4}$ .**

Therefore, taking into account the fact that the rate of restructuring reached after a given long period decreases very abruptly, it appears justified to consider this abruptness as revealing the system *metastability* as $\phi$ approaches $\phi_m$. Fig.3 shows evidence for this metastability. Define the *rate of restructuring* as the ratio $r(\phi) = 100*(S_{fin} - S_{init})/(1 - S_{init})$, where $S_{init} = S(t=0)$ and $S_{fin} = S(t=t_{fin})$. Figure 3 shows that when $\phi$ is increased above $\phi_{c2}$, the "final" value (for $\kappa_{A0} t_{fin} >> 1$) is increasingly far from $S = 1$, which is easily reached in the fluid phase, when $\phi < \phi_{c2}$.

## 4 - Evolution of $\eta(t)$ at constant shear rate: structuring and destructuring.

At constant shear rate, the viscosity always exhibits a plateau as $t \rightarrow \infty$, so the divergence of the viscosity and the bifurcation disappear. Using in eq.(6) a kinetic constant $\kappa_D$ which depends linearly on shear rate:

$$\kappa_D = k_C \, \dot{\gamma} \qquad (15)$$

defines the simplest NLS-model. Thus, at constant shear rate, this model does not predict either aging (in the sense used above) or, in consequence, rejuvenation. Moreover, for a given initial structure $S_{init}$ associated with an initial viscosity $\eta_{init}$, there is a critical value $\dot{\gamma}_K$

$$\dot{\gamma}_K = (\kappa_{FO}/\eta_{init})(1 - S_{init})/S_{init} \qquad (16)$$

which cancels the RHS side of eq.(6). This value leads to $S = constant$, thus keeping its initial value $S_{init}$, that is a constant viscosity $\eta = \eta_{init}$ from $t = 0$ onwards.

If $\dot{\gamma} > \dot{\gamma}_K$, one has $dS/dt < 0$, *i.e.* destructuring, which causes a viscosity decrease. Conversely, $dS/dt > 0$ if $\dot{\gamma} < \dot{\gamma}_K$, so the viscosity increases[3], due to structuring.

Fig.4 illustrates these results (with $\dot{\gamma}_K = 3.92$ for $S_{init} = 0.4$). It seems impossible to speak about aging in the case of this structuring and, *a fortiori*, about rejuvenation in the case of destructuring. These viscosity changes are only due to thixotropy.

---

[3] For $S_{init} = 0$ (*i.e.* if the agitation completely breaks down the structure), $S$ and therefore $\eta$ should increase, starting from its minimal value $\eta_\infty$, eq.(4).

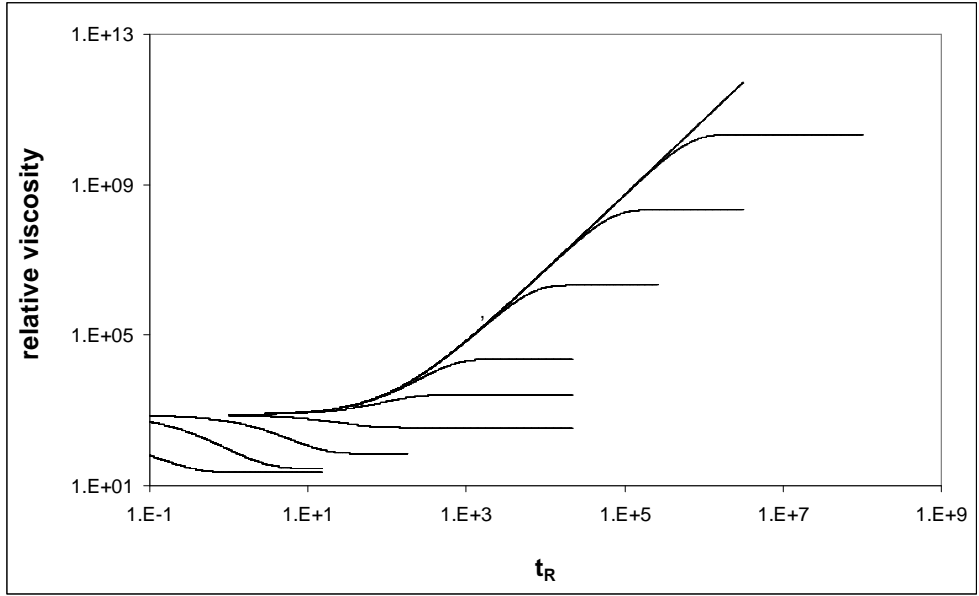

**Fig.4 –Relative viscosity vs. reduced time** $t_R = \kappa_{A0} t$ **under** $\dot{\gamma} = const.$

Values of $\kappa_D (s^{-1}) = kc \, \dot{\gamma}$   (growing from top to bottom) : 0 ; $10^{-7}$ ; $10^{-5}$ ; $10^{-3}$ ; $10^{-1}$ ; 1 ; 10 ; $10^2$ ; $10^3$ ; $10^4$

*The critical value which gives* $\eta = const$ *is* $\kappa_{DK} = 1.96$

($\phi = 0.50$ ; $k_C = 0.5$, $S_{init} = 0.4$ ; same values for other parameters as for Fig.1)

## 5 - Evolution of $\eta(t)$ under constant stress: aging and rejuvenation.

### 5.1- The NLS-model under constant stress.

Using the structural viscosity defined above, $\eta(t) = \eta[S(t)]$, the instantaneous shear rate under a constant shear stress is defined as $\dot{\gamma}(t) = \sigma / \eta(t)$, leading, with eq.(15), to a time dependent kinetic constant for destructuring, $\kappa_D = k_C \, \sigma / \eta(t)$.

Using the relative viscosity $\eta_R(t) = \eta(t) / \eta_F$ and the critical stress $\sigma_c = \kappa_{A0} \, \eta_F / k_C$ in eq.(6) gives:

$$\frac{dS}{dt} = \frac{\kappa_{A0}}{\eta_R(t)} \Big[ 1 - ( 1 + \sigma / \sigma_c ) S(t) \Big] \tag{17}$$

Under steady conditions, ($dS/dt=0$), the solution of eq.(17) is $S_{eq} = (1 + \sigma / \sigma_c)^{-1}$. From eqs.(1–5), the viscosity can be written as $\eta = \eta_F[1 - (1 - \chi) S_{eq}]^{-2}$ where $\chi = (1 - \phi / \phi_0) / (1 - \phi / \phi_m)$ is a rheological index [Quemada, 1998]. The steady viscosity is thus given by:

$$\eta_R = \eta_{R\infty} \left[ \frac{1 + \sigma / \sigma_c}{\chi + \sigma / \sigma_c} \right]^2 \tag{18}$$

This equation describes shear thinning behaviour when $\chi < 1$ and plastic behaviour when $\chi < 0$. In the latter case, the relative viscosity is given by:

$$\eta_R = \eta_{R\infty} \left[ \frac{\sigma + \sigma_c}{\sigma - \sigma_y} \right]^2 \qquad (19)$$

where $\sigma_y$ is the yield stress that depends on $\phi$ as:

$$\sigma_y = \sigma_c \left[ \frac{\phi/\phi_0 - 1}{1 - \phi/\phi_\infty} \right] \qquad (20)$$

### 5.2 – Aging and rejuvenation under constant shear stress

With $\sigma = constant$, the function $u = \eta_R^{1/2} = (1 - \phi_{eff}/\phi_m)^{-1}$ should verify the following equation, deduced from eq.(17):

$$\frac{du}{\beta + u^{-1}} = (1 + \sigma/\sigma_c)\,\kappa_{A0}\,dt \qquad (21)$$

with

$$\beta = \frac{\alpha(1 - \sigma/\sigma_B)}{1 + \sigma/\sigma_c} \qquad (22)$$

and

$$\sigma_B = \sigma_c \frac{\phi/\phi_{c2} - 1}{1 - \phi/\phi_m} \qquad (23)$$

$\alpha$ is defined by eq.(9).

As for $\alpha = 0$ in eq.(10), a bifurcation will be observed for $\beta = 0$ in eq.(21), *i.e.* when $\sigma$ reaches the critical stress $\sigma_B$ given by eq.(23). Note that this bifurcation only exists ($\sigma_B \geq 0$) if $\phi_{c2} \leq \phi \leq \phi_m$. Moreover, note that eq.(21) reduces to eq.(10) if $\alpha$ and $\kappa_{A0}$ are used in place of $\beta$ and $\kappa_{A0}(1 + A\sigma)$, in agreement with the limits at $\sigma = 0$. It is thus possible to deduce directly the solutions of eq.(21) at $\kappa_{A0}\,t \gg 1$ from the expressions for $u$ that led to eqs.(11−13). Therefore, at long times the viscosity can behave in three ways:

**a/** if $\sigma > \sigma_B$

$$\eta(t \rightarrow \infty) \approx \frac{\eta_F}{\alpha^2} \left[ \frac{\sigma / \sigma_c + 1}{\sigma / \sigma_B - 1} \right]^2 [\, 1 - exp(-B\kappa_{A0}t\,)] \tag{24}$$

where $B = \alpha^2 \, (\sigma/\sigma_B - 1)^2 \, (1 + \sigma/\sigma_c)^{-1}$

**b/** if $\sigma < \sigma_B$

$$\eta(t \rightarrow \infty) \approx \eta_F \, \alpha^2 (1 - \sigma/\sigma_B)^2 \, \kappa_{A0}^2 \, t^2 \tag{25}$$

**c/** if $\sigma = \sigma_B$

$$\eta = \eta_{init} + 2\eta_F (1 + \sigma/\sigma_c)\kappa_{A0}t \tag{26}$$

If $\sigma$ is varied, at constant volume fraction, on both sides of the bifurcation $\sigma = \sigma_B$, one thus recovers behaviour similar to that observed at $\sigma = 0$ as $\phi$ varied on both sides of the limit $\phi = \phi_{c2}$. In particular, if $\sigma \lesssim \sigma_B$, the absence of any limit for $\eta(t)$ will be associated with aging. On the contrary, if $\sigma > \sigma_B$, the existence of a finite value for the limit $\eta(t \rightarrow \infty)$ leads to a steady equilibrium that will be discussed below (section 6).

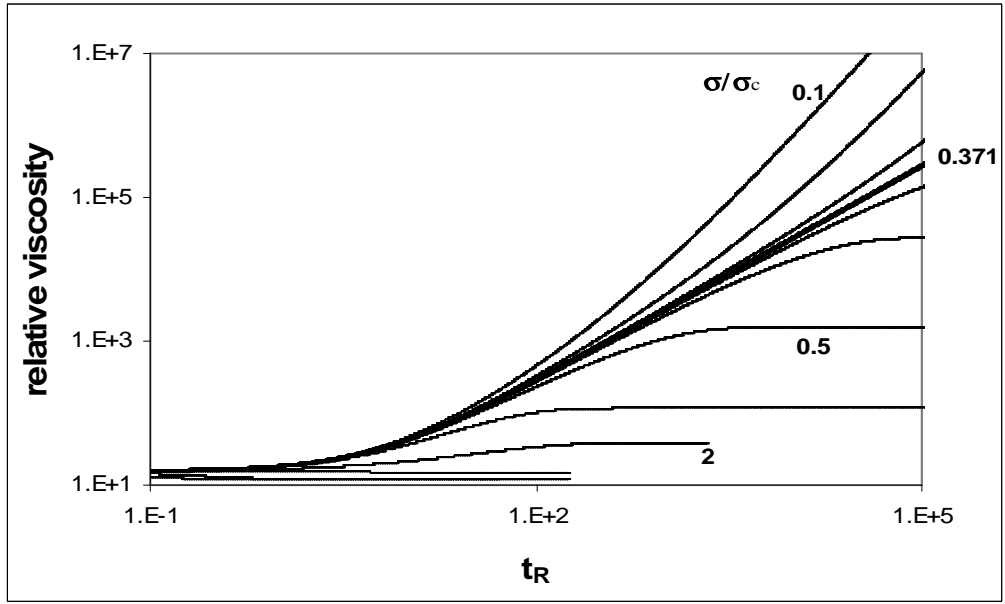

**Fig.5- Relative viscosity vs. reduced time** $t_R = \kappa_{A0}t$ .
**The bifurcation is observed at** $\sigma = \sigma_B$ **as the reduced stress** $\sigma /\sigma_c$ **is varied.**
Reduced stress values (from top to bottom) : 0.1 ; 0.3 ; 0.36 ; 0.371 ; 0.38 ; 0.4 ; 0.5 ; 1 ; 2 ; 10 ; 50 ; 100.
(Model parameters: $\phi = 0.45$; $\eta_F = 1$ mPa.s; $\sigma_c = 10$ Pa ; $\phi = 0.637$ ; $S_{init} = 0.1$ ; $\kappa_{A0} = 10^3$ s$^{-1}$).

Fig.5 shows evolution of the relative viscosity, $\eta_R$, as a function of the dimensionless time, $t_R = \kappa_{A0}t$, for different values of the reduced stress, $\sigma_R = \sigma /\sigma_c$, with $\phi_{c2} \lesssim \phi \lesssim \phi_m$. The lower limit is natural, since even at zero stress, there is insufficient structure for a bifurcation to occur. Fig.5 shows:

(i) steady viscosity plateaus, defined by eq.(24) as $\sigma > \sigma_B$;

(ii) the bifurcation line (with slope 1) at $\sigma = \sigma_B$,

(iii) viscosity divergences (asymptotes with slope 2) for $\sigma < \sigma_B$

However, a stress, $\sigma_K$, exists which cancels the RHS of eq.(17), so the system remains in its initial state ($S = S_{init}$). It is given by:

$$\sigma_K = \sigma_c (1 - S_{init})/S_{init} \tag{27}$$

For any stress in the range $\sigma_B \leq \sigma < \sigma_K$, increasing the viscosity from its initial value corresponds to restructuring ($dS/dt > 0$). For $\sigma > \sigma_K$, decreasing the viscosity from its initial value corresponds, in contrast, to destructuring ($dS/dt < 0$). This is shown in Fig.6 for $\sigma_K /\sigma_c = 9$ and $S_{init} = 0.1$. It seems difficult to argue that this destructuring, due to thixotropy, could be considered as *rejuvenation.* So, as in the case of evolution at rest in the fluid domain, discussed in section 3.2, it is impossible here to consider restructuring, also due to thixotropy, as an aging process.

On the other hand, for $\sigma \lesssim \sigma_B$, the unbounded viscosity increase, as well its slowing down (with a characteristic time $\approx$ system age), does seem to correspond to aging. Nevertheless, an open question remains: what it will happen if, after aging under a given stress $\sigma_I < \sigma_B$, applied up to a time $t_0$, a higher stress, $\sigma_2 > \sigma_I$, is applied from $t = t_0$? If destructuring occurred, at least at the beginning, should it be called rejuvenation? This point is discussed in section 5.4 .

Fig.6 shows the high shear stress behaviour, which is invisible in Fig.5, due to the scaling. On a linear scale, relative viscosity clearly decreases for $\sigma > \sigma_K$ (= $9\sigma_C$ here), illustrating the shear-induced destructuring.

Fig.6 shows a result comparable to that described in a recent article [**Coussot *et al*, 2002a**]. However, note that their model led to predictions which differ from those presented here. In this article, the stress value for bifurcation is defined by the structuring-destructuring equilibrium –hence it is identical to $\sigma_K$ in eq.(27). Therefore, it cannot be considered as an intrinsic characteristic of the material, since its value depends on the initial structure.

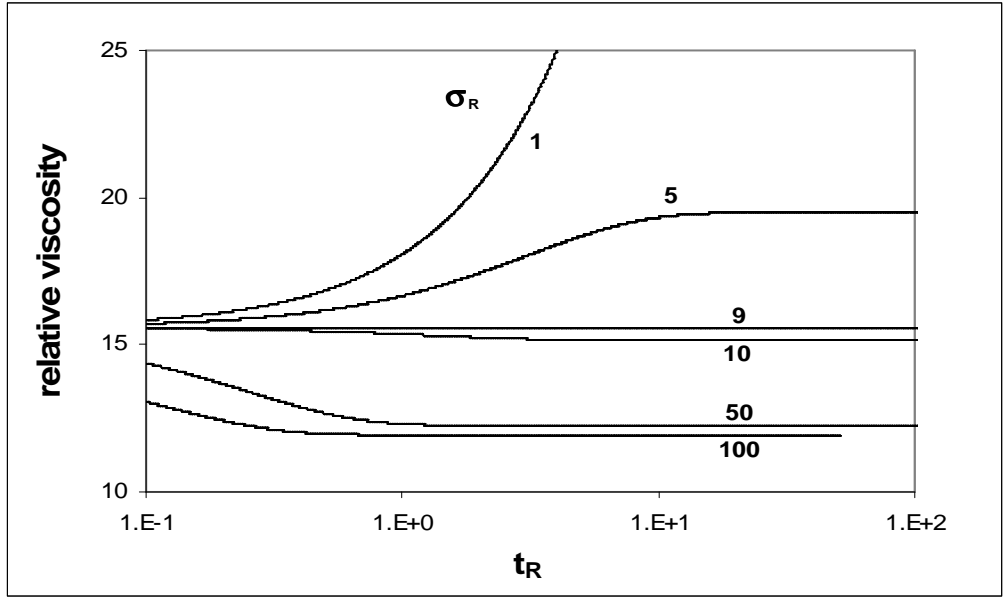

*Fig.6 – Idem Fig.5, but at high stresses .*
*"Rejuvenation" is observed for $\sigma_R > \sigma_{RK} = 9$ (see text).*
(Parameter values: *idem* Fig.5)

The corresponding constant viscosity $\eta = \eta_B$, thus separates the region of destructuring (where $\eta$ decreases towards a finite limit, $\eta < \eta_B$ ) from that of structuring (where $\eta$ increases without limit, $\eta \to \infty$ ). On the contrary, Fig.6 well illustrates a characteristic of thixotropic fluids: the viscosity increases towards a finite value when the stress applied to a system at equilibrium is decreased. The resulting increase of $\eta$ is due to the system restructuring, as shown in Fig.6, for $\sigma_R = 5$.

### 5.3 - Effects of variations of initial structure.

Figs.7A & 7B illustrate the effects of varying the initial structure, characterised by $S_{init}$, for $\sigma \leq \sigma_B$ and $\sigma > \sigma_B$ , respectively.

In the first case, the structural variable is bounded by the upper limit of $\phi_{eff}$ in the paste domain, $\phi_m$. From eq.(2), this corresponds to the maximum value $S_m = (\phi_m /\phi -1)/C$. As $S_{init}$ increases, both $\phi_{eff}$ and $\eta$ increase. However, as $dS/dt \approx 0$, there is a quasi-stationary viscosity[4] along a higher and higher $\eta$-plateau of longer and longer duration, before the $\eta$-divergence as $t^2$ occurs. This is shown on Fig.7A where, as $S_{init}$ is approaching $S_m$ , all $\eta$ - plateaus $\to \infty$ .

---

[4] This corresponds to the abrupt fall of the restructuring rate shown in Fig.3 .

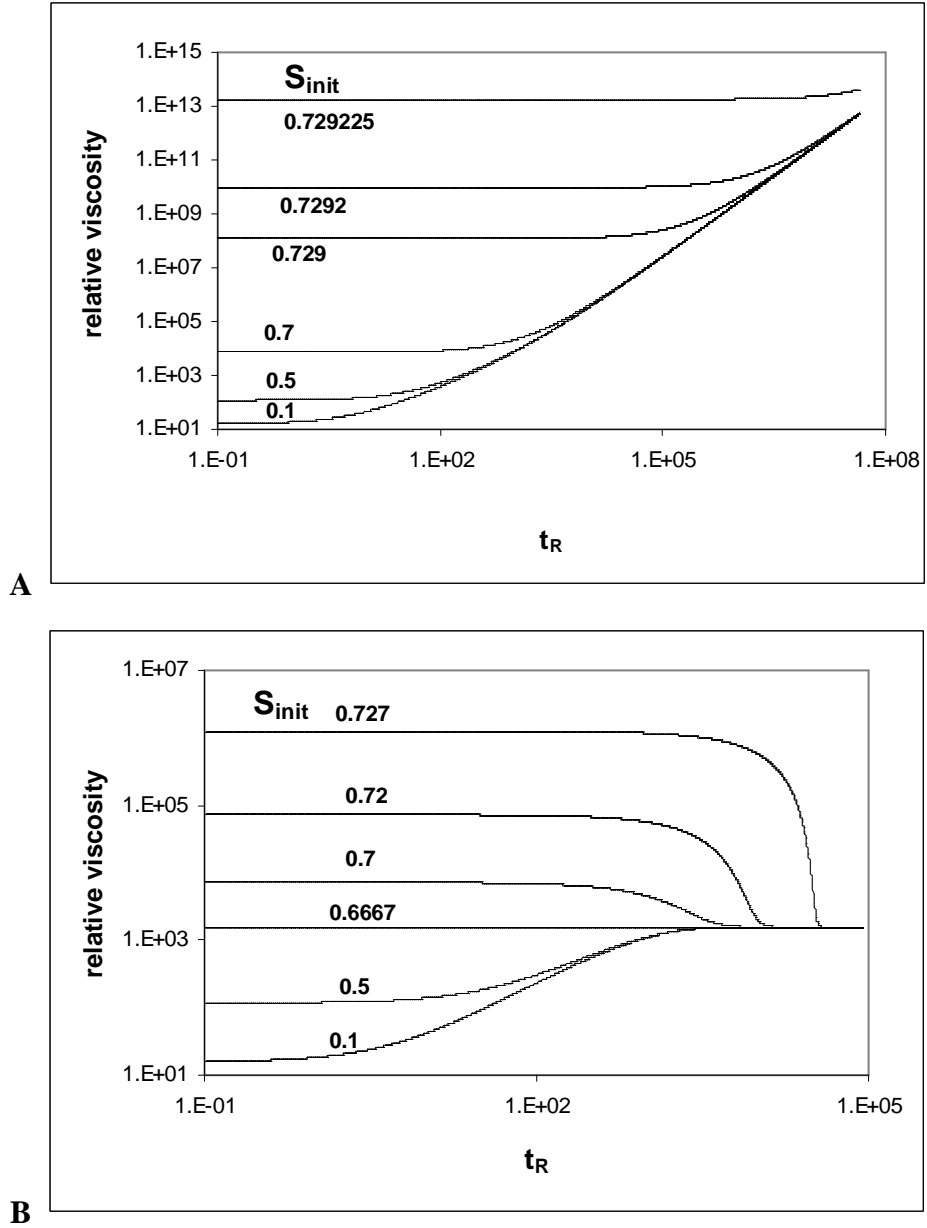

***Fig.7** - Evolution of η(t) resulting from applying a constant stress,
following rest periods of increasing lengths
(thus leading to growth of $S_{init}$ , however bounded by $S_m$ )*
**(A)** $\sigma < \sigma_B$ , with $S_m = 0.729$ and $S_{eq} = 0.833$       **(B)** $\sigma > \sigma_B$ , with $S_m = 0.729$ and $S_{eq} = 0.667$
*(Model parameters: $\phi = 0.45$ ; $\eta_F = 1\ mPa.s$ ; $\sigma_c = 2\ Pa$ ; $\varphi = 0.637$ ; $S_{init} = 0.4$ ; $\kappa_{A0} = 10^3\ s^{-1}$ ).*

In the second case, $\sigma > \sigma_B$, the equilibrium structure is always accessible since $S_{eq} < S_m$ , leading to a finite limit for $\eta(t \rightarrow \infty)$. As $S_{init}$ is increased, there is first restructuring, if $S_{init} < S_{eq}$ , then destructuring if $S_{init} > S_{eq}$ (with obviously $\eta = const$ for $S_{init} = S_{eq}$). As in the case $\sigma \leq \sigma_B$ , $\eta$ increases more and more as $S_{init}$ approaches $S_m$ , giving again a higher and higher $\eta$-plateau of longer and longer duration and a quasi-stationary state with $dS/dt \approx 0$, even though the equilibrium structure, $S_{eq} = (1 + \sigma / \sigma_c)^{-1}$, was not reached. Fig.7B shows these properties, where the trends are similar to those observed experimentally in bentonite suspensions for increasing rest times (cf. [**Coussot et al, 2002b, Fig.5**]). The results in Fig.7A & B show that, under given stress $\sigma_A$ , it is not possible to obtain the bifurcation by modifying $S_{init}$ : either $\eta$ diverges, if $\sigma_A \leq \sigma_B$ , or $\eta$ remains finite if $\sigma_I > \sigma_B$ . In this model, the stress bifurcation is *intrinsic*, *independent of initial conditions*, in contrast to the predictions of Coussot and co-workers.

In both cases, $\sigma \leq \sigma_B$ or $\sigma > \sigma_B$ , ranges of quasi-steady viscosity can be considered as revealing a system metastability which grows with $\phi$ . Moreover, the presence of such viscosity plateaus might explain : i) the time-dependence of the yield stress [**Nguyen & Boger, 1992**], which is very frequently observed in hysteresis cycles and ii) much of the difficulty in the experimental determination of the yield stress (see [**Cheng, 1986**] for instance).

### 5.4 – Rejuvenation under constant stress after aging

When a stress acts on a system which has aged, there is a return towards the previous state. It is natural to call this return *rejuvenation*. On the other hand, fluids and soft pastes evolve towards an equilibrium state, so such a process is quite foreseeable in the framework of a thixotropic model (remember that thixotropic systems are reversible). In this case, it seems clear that the term "rejuvenation" is inappropriate. However, it is much less clear for hard pastes, since they do not evolve towards equilibrium, so they may lose reversibility. The following discussion will demonstrate that the NLS-model can predict behaviour which can, to some extent, be considered as rejuvenation.

If a system with $\phi > \phi_{c2}$ is at rest[5] for a time $t_0$ , it ages with a viscosity which grows as $t^2$ , in agreement with eq.(12), as displayed on Fig.8, curve(a). At $t = t_0$ , one applies a stress $\sigma_m$ which is maintained for $t > t_0$ .

Several cases are possible, according to the $\sigma_m$-value in comparison with $\sigma_B$ on the one hand and, on the other hand, a value $\sigma_K$ still defined by eq.(27), but with $S_{init} = S(t_0)$. Fig.8 illustrates three cases, for values of $\sigma_m$ (*m = 1, 2, 3*):

(i) for $\sigma_I$ such that $\sigma_I > \sigma_K > \sigma_B$ , destructuring occurs at $t > t_0$ . The viscosity decreases to the steady value corresponding to the equilibrium structure reached under $\sigma_I$. Aging thus stops at $t_0$ and it seems impossible to speak about rejuvenation after $t = t_0$. This is shown in curve (b) of Fig.8, for values of the relative stress $\sigma_R = \sigma / \sigma_c$ : $\sigma_{RI} = 1.5$ ; $\sigma_{RK} = 1.35$ and $\sigma_{RB} = 1.08$.

---

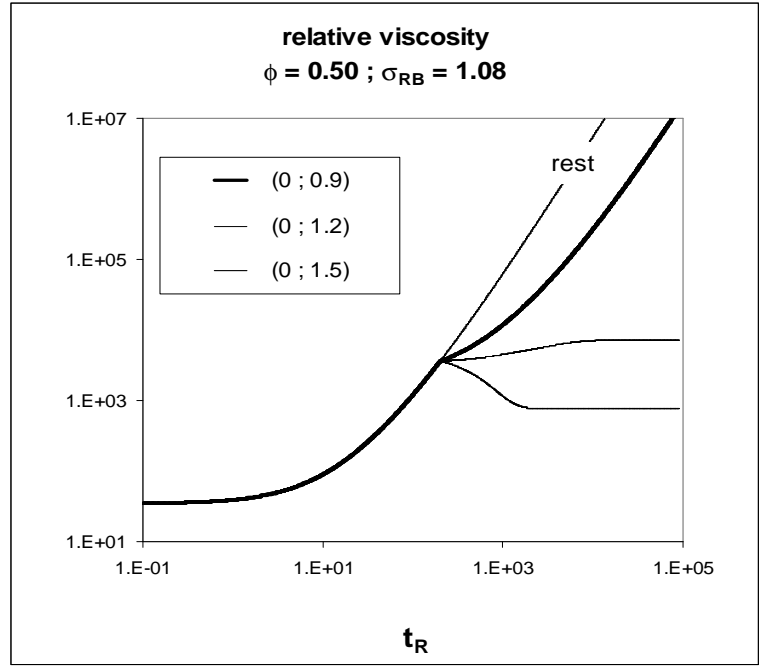

*Fig.8 - Relative viscosity vs. reduced time $t_R = \kappa_{A0} t$ .*

*Effect of applying a reduced stress $\sigma_A /\sigma_c = \sigma_{RA}$ , following a rest period ($\sigma_{R0} = 0$ ; duration $t_{R0}$).*

*Curve Index*: ( $\sigma_{R0}$ ; $\sigma_{Rm}$ ) means that $\sigma_R = \sigma_{R0}$ for $0 < t_R < t_{R0}$ and $\sigma_R = \sigma_{Rm}$ for $t_R > t_{R0}$ .

"rest" corresponds to (0 ; 0) -- Bifurcation appears for $\sigma_R = \sigma_{RB} = 1.08$ , value in between $\sigma_{R1}$ and $\sigma_{R2}$. -

Duration: $t_{R0} = 200$. (same values of other NLS-parameters as for Fig.4)

ii) for $\sigma_2$ such that $\sigma_k > \sigma_2 > \sigma_B$ , structuring continues beyond $t = t_0$, however without possibility to strictly speak about aging since, under $\sigma_2 = const,$ the viscosity grows towards a *finite value*. This means again that aging stops at $t_0$ (see Fig.8 curve (c), for $\sigma_{R2} = 1.2$).

iii) in contrast, for $\sigma_3 > \sigma_B$ , structuring continues beyond $t_0$, with a viscosity which diverges as $t^2$. Hence, aging now continues as shown in Fig.8 curve(d) for $\sigma_{R3} = 0.9$ .

Nevertheless, in the latter case, this divergence is delayed in comparison with what would be observed if the system remained at rest after $t = t_0$ . In other words, the viscosity at each time $t_1 > t_0$ is lower than that it would be at the same time $t_1$ if it had remained at rest. This *slowing down of aging* could be termed *rejuvenation*.

It is worth noting one rather general characteristic : for a given system, the magnitude of applied stress at $t = t_0$, compared to that of $\sigma_B$, given by eq.(23), fixes the future behaviour of the system, whatever the duration of $t_0$ .

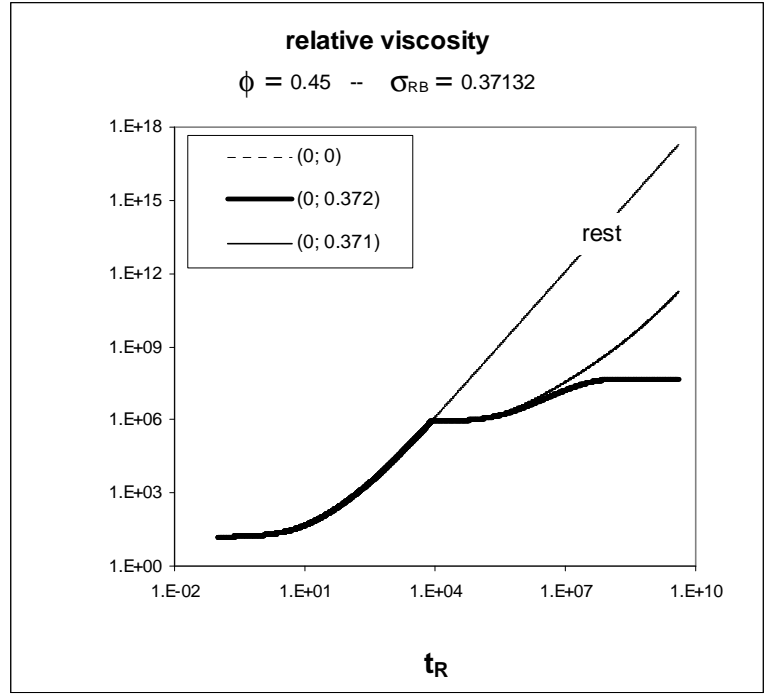

*Fig.9* - *Evidence of the bifurcation at* $\sigma_R = \sigma_{RB}$

*under applying a stress after a rest period (duration* $t_{R0} = 8.10^3$).

*Rejuvenation under shear rate is associated to variations under* $\sigma_R < \sigma_{RB}$

(same SNL-parameter values than for Fig.4).

Fig.9 illustrates this point. It shows one evolution from each side of the bifurcation. In comparison with aging at rest, the evolution for $\sigma_R$, which is slightly lower than $\sigma_{RB}$, shows an extremely large "rejuvenation". To recover the same state of aging (choose the viscosity to define the age, for instance the value $\eta_R \approx 10^{10}$ reached after a rest duration $t_{R0} \approx 10^6$), one must wait for a very large time ($t_R \approx 10^9$) for the system submitted to $\sigma_R = 0.371$ at $t = t_{R0}$ to reach this state. Such a delay could be taken as a measure of the rejuvenation, still keeping this term to mean a *slowing down of aging*.

## 6 – Steady viscosity: shear-thinning and plastic behaviour

For $\sigma > \sigma_B$ , eq.(24) gives the variations $\eta(t)$. Under steady conditions ($t \rightarrow \infty$), one recovers the behaviour of a fluid with a yield stress $\sigma_B$ defined by eq.(23). The steady viscosity is obtained from eq.(24):

$$\eta_{eq} = \frac{\eta_F}{(1 - \phi / \phi_m)^2} \left[ \frac{\sigma + \sigma_c}{\sigma - \sigma_B} \right]^2 \qquad \text{for } \sigma > \sigma_B \qquad (28)$$

In recent work, this plastic behaviour, usually observed experimentally, has been very often represented by an empirical relation like the Herschell-Bulkley (HB) law

$$\sigma = \sigma_y + K\,\dot{\gamma}^{\,m} \qquad\qquad \text{for } \sigma > \sigma_y \qquad\qquad (29)$$

In order to compare eqs.(28) and (29), they must be written as $\dot{\gamma} = f(\sigma)$. Eq.(28) becomes

$$\dot{\gamma} = \frac{\sigma}{\eta_\infty}\left(\frac{\sigma - \sigma_B}{\sigma + \sigma_c}\right)^2 \qquad\qquad \text{for } \sigma > \sigma_B \qquad\qquad (30)$$

$$\dot{\gamma} = 0 \qquad\qquad \text{for } \sigma \le \sigma_B$$

using in eq.(30), the "high shear" viscosity limit $\eta_\infty$ at $\sigma \gg \sigma_B$ and $\sigma \gg \sigma_c$, given by eq.(4). Eq.(29) can be expressed as:

$$\dot{\gamma} = \left(\frac{\sigma - \sigma_y}{K}\right)^{1/m} \qquad\qquad \text{for } \sigma > \sigma_y \qquad\qquad (31)$$

$$\dot{\gamma} = 0 \qquad\qquad \text{for } \sigma \le \sigma_y$$

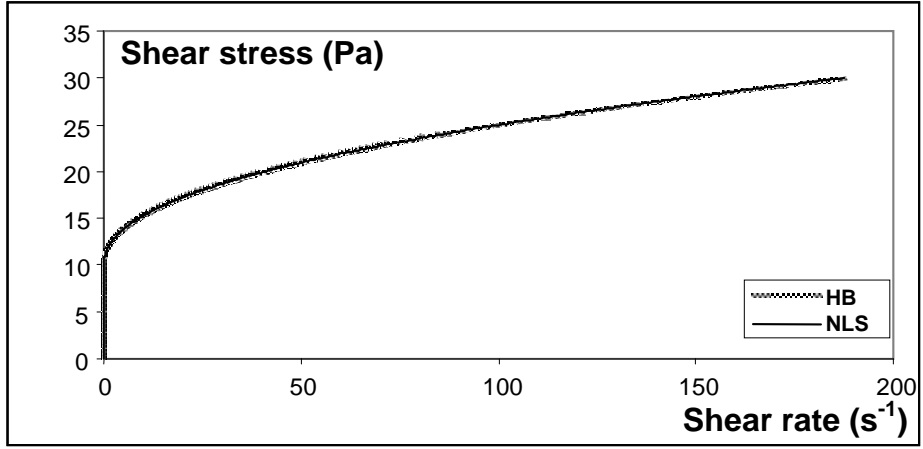

*Fig.10  - Fitting HB-model to NLS-model curves*

Parameters: NLS: $\eta_\infty = 10\ mPa$ ; $\sigma_c = 50\ Pa$ ; $\sigma_B = 10\ Pa$

HB: $K = 1.806\ Pa.s^m$ ; $m = 0.458$ ; $\sigma_y = 10.1\ Pa$

Despite the very different forms of eqs.(30) and (31), it is always possible to adjust the HB-model parameters $K$ and $\sigma_y$ to fit the variation described by eq.(28), with a fair determination of the yield stress. Clearly, the determination will be better if the fit is confined to data measured at low shear rates.

Fig.10 illustrates a fit to the flow curve calculated using the same parameters as were used for Fig.1. The agreement is very satisfactory, especially near the yield stress[6]. Obviously, this fit will not work at high shear rates, since the asymptotic behaviour of eq.(31) is only identical to the linear one of eq.(30) in the case of Bingham fluids ($m=1$). A large number of theoretical and experimental studies have shown $0.5 < m < 0.9$, so using the NLS-model should be a step forward, compared with using a simple model, such as Herschell-Bulkley, which is the usual procedure in the literature.

Finally, note that eqs.(28) and (23) are quite consistent with eqs.(19) and (20), if one takes $\phi_0 \equiv \phi_{c2} = \varphi \, \phi_m$ , $\phi_\infty \equiv \phi_m$ . See eq.(5). These values correspond to steady limits of $\phi_{eff}$ $= \phi \, (1+CS)$ at very low and very high shear rate : as $\phi \geq \phi_{c2}$ , one has the following limits at $t \rightarrow \infty$ :

   (i) $S = 1$ for $\sigma/\sigma_c << 1$, giving $\phi_{eff} = \phi \, / \varphi$ , hence $\phi_{eff}/\phi_m \equiv \phi \, / \phi_{c2}$ ;

   (ii) $S = 0$ for $\sigma/\sigma_c >> 1$, giving $\phi_{eff} \equiv \phi$ .

Thus, the limiting packing fractions $\phi_0$ at $\sigma = 0$ and $\phi_\infty$ as $\sigma \rightarrow \infty$ are identical to $\phi_{c2}$ and $\phi_m$, respectively.

## 7 - Comparison with experimental data.

The time-dependent behaviour of colloidal suspensions of synthetic Laponite at constant shear rate has recently been interpreted in terms of aging and rejuvenation [**Abou et al, 2003**]. Here the aim is to show that this data can be modelled using the ("purely thixotropic") NLS-model. This approach needs to choose the values of volume fraction $\phi$ and kinetic constants $k_C$ and $\kappa_{A0}$ involved in eqs.(15) and (6). Such "modelling" of experimental data can be performed[7] by only changing the values of $\phi$ and the pair ($k_C$ ; $\kappa_{A0}$) that will be respectively associated to parameters which characterize each measured sample, *i.e.* weight concentration, $c_m$, and ionic strength, $I$ . Although it seems plausible that $I$ influences the kinetic processes, close association of $I$ with $k_C$ and $\kappa_{A0}$ would require complex theoretical work, which is not attempted here. Thus, here the correspondence is only phenomenological. In figure captions, it will be noted:

   $[\phi \, ; \, k_C \, ; \, \kappa_{A0} \, (s^{-1})] \leftrightarrow [c_m \, \% \, ; \, I \, mM])$.

---

[6] For instance (Fig.10), the relative error in $\sigma_y$ is less than 2 % for $0 < \dot{\gamma} < 200 \, s^{-1}$ .

[7] keeping the other NLS parameters unchanged : $\phi_m = 0.637; \, \eta_F = 1 \, mPa.s$ ; $\varphi = 0.637$ however with $S_{init} = 0$ .

Figs.11, 12, 13 and 14 are comparable to experimental data [**Abou** *et al*, **2000** ; Figs.2, 4, 5 and 6, respectively[8]]. The figures show a fair correspondence between $\phi$ and $c_m$ and, the pair $(k_C$ ; $\kappa_{A0})$ and $I$ . Naturally, free values for $\phi$ , $k_C$ and $\kappa_{A0}$ have been used, although imposed to be common for all data having same sample characteristics $(c_m=1.5\%; I=7.10^{-3})$, hence for Figs.11(a), 12 and 13.

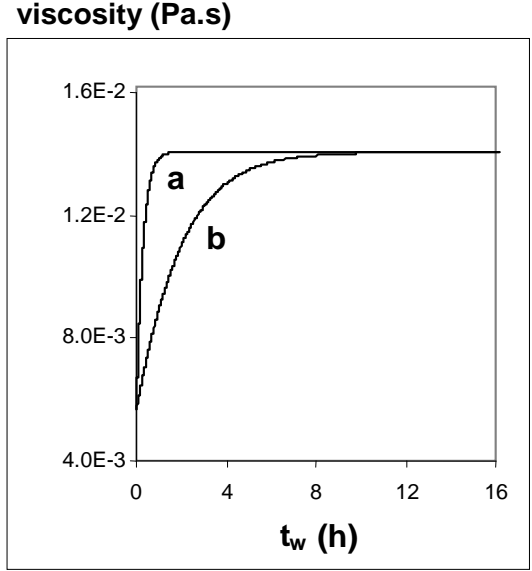

*Fig.11 – Restructuring under* $\dot{\gamma} = 500\ s^{-1}$
**(a)** [0.35 ; 2.10⁻³; 20] ↔ [1.5 ; 7.10⁻³] ;
**(b)** [0.35 ; 3.10⁻⁴; 3] ↔ [1.5 ; 5.10⁻³]

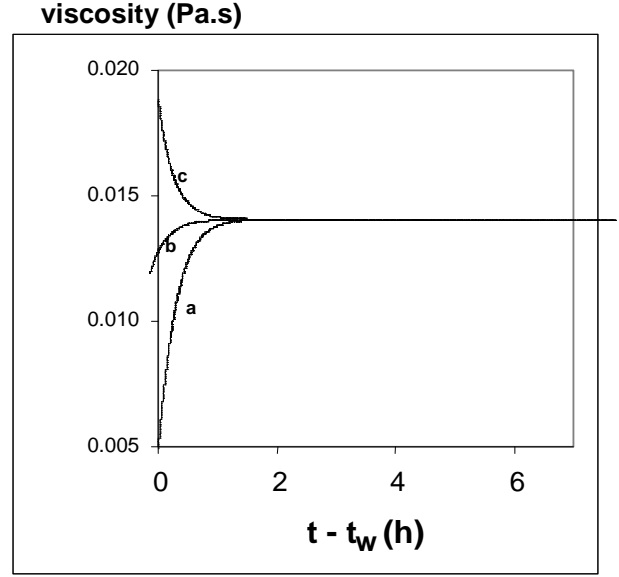

*Fig.12 – Restructuring or destructuring*
*under* $\dot{\gamma} = 500\ s^{-1}$ *starting from states first submitted to*
*restructuring during periods of rest during*
*(a):* $t_w = 0$ *min;* *(b):* $t_w = 20$ *min;* *(c):* $t_w = 40$ *min.*
*With* [0.35 ; 2.10⁻³; 20] ↔ [1.5 ; 7.10⁻³]

Comparison of Fig.11(b) with Fig.11(a) shows that slightly decreasing $I$ leads to a change in $k_C$ and $\kappa_{A0}$ , however their ratio $(k_C / \kappa_{A0} = 10^{-4})$ is unchanged, which is compatible with an increase of effective particle radius.

Fig.12[9] shows viscosity changes towards equilibrium states after different periods at rest. Restructuring occurs in curves (a) and (b)[10], while destructuring occurs in curve (c), since the initial state is more structured than at equilibrium. As a matter of fact, after complete destructuring $(S_{init} = 0)$, "aging dynamics" (in fact, restructuring) under $\dot{\gamma} = 0$ , $S(t)$ reaches the values $S(20\ min)= 0.546$ and $S(40\ min)= 0.703$ . These values are the $S_{init}$ ones, required to calculate evolutions under $\dot{\gamma} = 500\ s^{-1}$, shown, respectively, in Fig.12, (b) & (c).

---

[8] Figs.1 and 3 from [**Abou** *et al*, **2003**] were not used, as they concern complex viscosity, which is not considered here.
[9] with the same parameter values as in Fig.11(a).
[10] curve (a) is identical to curve (a) of Fig.11

In the case of Fig.13, using same parameter-values leads to plateaus at *100* and *50 s⁻¹*, lower than the measured ones. Note that this could obtained by assuming a slight dependence of $k_C$ on $\dot{\gamma}$ (*i.e.* discarding the hypothesis of linearity in $\dot{\gamma}$ for structure kinetics, assumed here for simplicity).

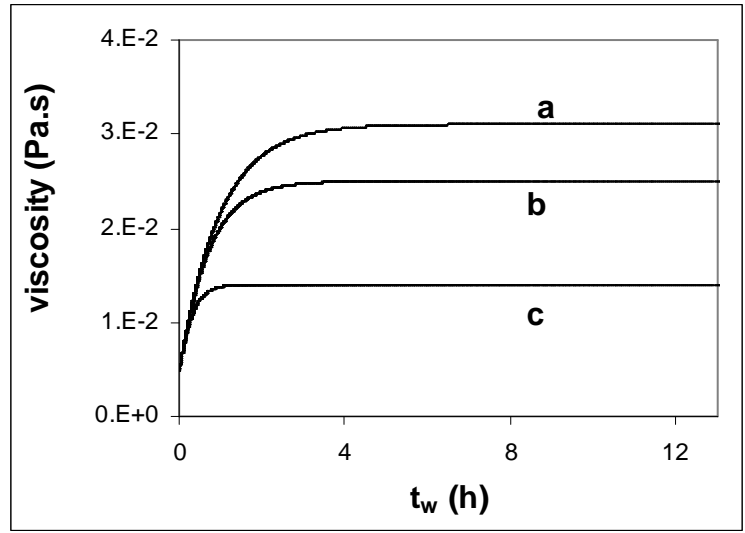

***Fig.13** – "Rejuvenation" under constant shear rates $\dot{\gamma}$ ( s⁻¹) = **(a): 50 ; (b): 100 ; (c): 500** . (see text).*
*(with [0.35 ; 2.10⁻³; 20] ↔ [1.5 ; 7.10⁻³])*

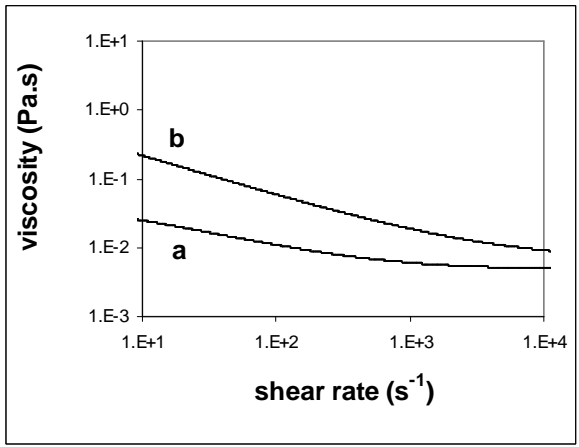

***Fig.14** – Steady viscosity vs $\dot{\gamma}$.*
*(Curve drawing limited to used experimental domain (10 < $\dot{\gamma}$ <10⁴)*
**(a) [0.35 ; 2.10⁻³; 20] ↔ [1.5 ; 7.10⁻³].**
**(b) [0.40 ; 4.10⁻³; 500] ↔ [3.7 ; 10⁴].**

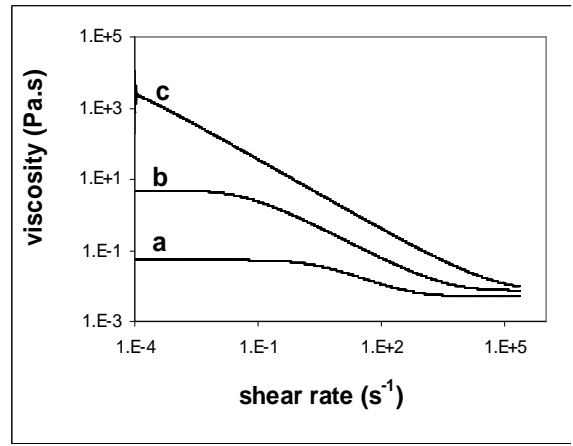

***Fig.15** – Steady viscosity vs $\dot{\gamma}$ .*
**(a) [0.35 ; 2.10⁻³; 20]**
**(b) [0.40 ; 4.10⁻³; 500]**
**(c) [0.4057 ; 4.10⁻³; 500].**

Finally, acceptable model predictions of steady behaviour are shown by curves (a) and (b) in Fig.14. However, variations of $\phi$ and $c_m$ occur in the right way in contrast with those of $(k_C; \kappa_{A0})$ and $I$. Such discrepancies could result, at least partially, from the difficulty of experimentally reaching a *true* steady state: the laponite samples under study require restructuring times of many hours.

Nevertheless, it is satisfying to see in Fig.15 that the same parameters used for Fig.14 lead to Newtonian plateau viscosities at very low and very high shear rates, as expected for a concentrated dispersion. Curves (a) and (b) in Fig.15 indeed represent the same curves as the corresponding ones on Fig.14, but over a larger range of shear rate. Moreover, curve (c) shows that the limit of plastic behaviour (with $\eta \to \infty$) is well recovered as $\phi_{eff} \to \phi_{c2}$ (here $\phi_{c2} = 0.40577$, keeping same values of $\phi_m$ and $\varphi$).

The relationship between a) the model values: $\phi$, $k_C$ and $\kappa_{A0}$ and b) the experimental values: weight concentration $c_m$ and ionic strength, $I$, have not been yet discussed. It will now be shown that use of the NLS-model is justified if the complex structure of laponite dispersions is taken into account.

First, it seems clear that the equivalent particles (the "Hard Spheres" of the NLS-model), *a priori* unidentified, are large, mesoscopic structures rather than at the platelet-scale[11]. Moreover, the $c_m$-values can be linked to $\phi$, the HS-volume fraction, by taking into account the structural characteristics of the system :

(i) the laponite Primary Particles (PP) studied in [**Abou** *et al*, **2003**], are disks with diameter *2a = 25 nm*, thickness *e = 1 nm* and specific mass $\rho = 2.5$ *g/cm³*

(ii) structures analysed by light, neutron and X ray scattering [**Pignon** *et al*, **1997**] and by cryofracture, TEM et SAXS [**Mourchid** *et al*, **1995**] as detailed below.

For *I = 10⁻³ M* and a weight fraction $c_m > 1.5$ %, [*i.e.* a *true* volume fraction $c_V = (c_m/\rho) > 0.006$], the system at rest consists of *structures at three length scales* : a network of *fractal clusters* (FC) with radius *R* and fractal dimension *D*, each of them made up of *dense aggregates* (DA) with radius *r*, themselves made up of sub-units of radius $a_P$ (cf. [**Pignon** *et al*, **1997, Fig.3**]). The latter could be considered as *oriented microdomains* (OM) in the form of platelets [**Mourchid** *et al*, **1995**], with diameter $2a_P$ and thickness $e_P$. This thickness is the worst defined parameter. Then, the effective volume fraction of OM is evaluated[12] from their hydrodynamic volume $\propto (a_P)^3$, leading to $\phi_{OM} = (4/3)(a_P/e_P)c_V$. If the DA-compactness is close to random close packing ($\varphi_{RCP} = 0.64$), the volume fractions of DA and FC are respectively $\phi_{DA} = \phi_{MO}/\varphi_{RCP}$ and $\phi_{FC} = \phi_{DA}(R/a_P)^{(3-D)}$.

---

[11] with interactions of the "soft sphere" type at low *I* [**Levitz** *et al*, **2000**]).

[12] A similar definition has already been used to model montmorillonite suspensions [**Baravian** *et al*, **2003**].

**Application to data** [**Abou** *et al*, **2003**] (with $c_m = 0.015$ *g/ml* , thus $c_V = 0.006$). For the following evaluation, take $2a_P = 30$ *nm* for platelets and $2R \approx 5$ *µm*, $D = 1.8$, $2r \approx 1$ *µm* for the mesoscopic structures observed at $I = 10^{-3}$ *M* [**Pignon** *et al*, **1997**]. Choosing $3 \le e_P \le 4$ *µm* , one obtains[13] $0.43 \ge \phi_{FC} \ge 0.32$ . Without considering that this result confirms the value $\phi = 0.35$ used for modelling, it supports identifying the FC as the model Hard Spheres, hence with a volume fraction $\phi >> c_V$ . Furthermore, if the value (= *23*) of the ratio $\phi/c_m$ was supposed to be maintained[14] as $c_m$ increases, the sol-gel transition would be observed at $c_m{}^* = \phi_G/23 = 0.025$ (*i.e.* $c_V{}^* = 0.01$). Three arguments suggest that this transition value is compatible with those deduced from experiments: (i) for the re-entrant transition line in the phase diagram [**Levitz** *et al*, **2000**] ; (ii) for the beginning of the viscosity divergence [**Baravian** *et al*, **2003**] ; (iii) for the change (*2 → 3*) of exponent *m* in the power law describing the $\phi$-dependent yield stress, $\sigma_y \propto \phi^{m}$ [**Pignon** *et al*, **1997**].

Obviously, an actual data fitting with the NLS-model could be obtained only from better knowledge of changes in laponite under shear. This could be achieved by measurements, both structural and rheological, under transient conditions in order to improve modelling of the structure kinetics.

## 7 – Conclusions and future prospects

In conclusion, both for systems at rest and under controlled stress, the NLS-model of thixotropy has been shown to give satisfactory modelling of aging and, to some extent, shear-induced rejuvenation. This result has been obtained through giving prominence to bifurcations in the time-evolution of viscosity.

At rest, different domains of volume fraction have been found, that depend on the initial structure, *i.e.* the previous history of the material. This has led to distinguish fluid and paste domains, the latter being divided into two states: hard and soft pastes.

At constant shear rate, the model predicts neither aging nor rejuvenation. Only shear-induced restructuring or destructuring occurs, depending on initial state of the material.

On the contrary, under constant shear stress, there is a viscosity bifurcation at a critical stress. The critical value has been found to be the plastic yield stress, which depends on the volume fraction. As the latter is usually described by the empirical Herschell-Buckley law with constant yield, some progress can be expected by using instead the present, physically based modelling.

---

[13] Larger values, $0.50 \ge \phi_{AF} \ge 0.37$ could be obtained using the size $2a_{eff} = 35$ *nm* observed in [**Pignon** *et al*, **1997**].

[14] A hypothesis that is valid for fractal mesostructures.

The importance of distinguishing aging and rejuvenation from simple structural changes, restructuring or destructuring, either due to thixotropy alone or also in the presence of shear has been emphasized. This need has been illustrated by a satisfactory comparison of model predictions with data on laponite submitted to shear rate steps and under steady conditions.

Finally, as the model works under all types of transient conditions, further testing can be done using stress relaxation and hysteresis cycles.

## Acknowledgments.

I thank A.Parker for fruitful discussions.